\documentclass{aa}  
\usepackage{bm}
\usepackage{mathrsfs}
\usepackage{subfigure}
\usepackage{graphicx}
\usepackage{color}
\usepackage{ulem}
\usepackage{txfonts}

\usepackage[colorlinks=true, citecolor=blue, linkcolor=blue]{hyperref}

\begin{document}

   \title{Studying the diffusion mechanism of cosmic-ray particles}


   \author{Ya-Wen Xiao\inst{1}\and
          Jian-Fu Zhang\inst{1}\fnmsep\inst{2}\and
          Siyao Xu\inst{3}
          }

   \institute{Department of Physics, Xiangtan University, Xiangtan, Hunan 411105, People's Republic of China \\
              \email{jfzhang@xtu.edu.cn}
         \and
             Key Laboratory of Stars and Interstellar Medium, Xiangtan University, Xiangtan 411105, China\\
         \and
             Institute for Advanced Study, 1 Einstein Drive, Princeton, NJ 08540, USA\\
             }


 
  \abstract
{More and more observations have indicated the existence of slow diffusion phenomena in astrophysical environments, such as around the supernova remnants and pulsar $\gamma$-ray halos, where the diffusion coefficient of cosmic rays (CRs) near the source region is significantly smaller than that far away from the source region. The inhomogeneous diffusion indicates the existence of multiple diffusion mechanisms.}
{Comparing the CR mirror diffusion with the scattering one, we aim to explore their diffusion characteristics in different magnetohydrodynamic (MHD) turbulence regimes and understand the effect of different MHD modes on mirror and scattering diffusion.}
{We perform numerical simulations with the test particle method. Within the global frame of reference, we first measure parallel and perpendicular CR diffusion and then determine the mean free path of CRs with varying energies.}
{Our main results demonstrate that: (1) CRs experience a transition from superdiffusion to normal diffusion; (2) mirror diffusion is more important than scattering diffusion in confining CRs; (3) CR diffusion strongly depends on the properties of MHD turbulence; and (4) magnetosonic and Alfv\'en modes dominate the parallel and perpendicular diffusion of CR particles, respectively.}
{The diffusion of CRs is a complex problem of mixing the mirror diffusion and scattering diffusion. The property of turbulent magnetic fields influences CR diffusion. The CR slow diffusion due to the presence of magnetic mirrors in turbulence has important implications for explaining observations near a CR source.}

   \keywords{cosmic rays --
                diffusion --
                magnetohydrodynamics --
                turbulence
               }

   \maketitle
%

\section{Introduction}
Cosmic rays (CRs), composed of high-energy charged particles, play an important role in astrophysics. Currently, the origin of Galactic CRs, particularly PeV CRs, is still an open issue (e.g., \citealt{Amato_Casanova_2021}). From an observational point of view, detecting the fluxes of secondary and primary CR particles can constrain the CR diffusion behavior \citep[e.g.,][]{Amato2014IJMPD}. Interestingly, \cite{Abeysekara2017Sci} found that the CR diffusion coefficient in TeV halos surrounding the Geminga and PSR B0656+14 is smaller by a factor of about 100 than the values constrained in the diffusion model of electron diffusion into the local interstellar medium (ISM). Studying the diffusion and acceleration of CRs in magnetized turbulent ISM is an effective way to understand the origin of the Galactic CRs. Meanwhile, it can help us to understand some key astrophysical processes such as the formation and evolution of galaxies and stars, the radiation of galaxy clusters, and diffuse $\gamma$-ray emissions (\citealt{Rodgers-Lee2020MNRAS, Semenov2021ApJ, Brunetti2007MNRAS, Krumholz2020MNRAS, Yan2012ApJ}).

Properly understanding the properties of the magnetohydrodynamic (MHD) turbulence is crucial for exploring the CR propagation processes. Traditionally, the slab/2D composite model of MHD turbulence was used to explain solar wind turbulence (e.g., \citealt{Bieber1988ApJ, Matthaeus1990JGR}). Significantly, the new paradigm of MHD turbulence theory (\citealt{GS95}, hereafter GS95) has advanced the study of the interaction of CRs with magnetized ISM (see \citealt{BL2019.book} for more details). Under the condition of the critical equilibrium, i.e., the time of the propagation of waves equal to the turbulence cascade time, GS95 theory proposed that the motion of turbulence both in perpendicular and parallel directions with respect to the magnetic field is different. The anisotropy of MHD turbulence has been verified by numerical simulations (\citealt{CL02, CL03}, hereafter CL02, CL03). By decomposing MHD turbulence into three MHD modes: Alfv\'en, slow, and fast modes, CL02 found that the incompressible Alfv\'en mode and compressible slow mode are anisotropic, and the compressible fast mode presents isotropy.  

Under the assumption of infinitesimal magnetic fluctuations, the quasi-linear theory (QLT) (\citealt{Jokipii1966ApJ}) can describe the transport of particles. The QLT analytically predicted the relationship of $\lambda_{\parallel} \propto \rho^{2-s}$ ($s$ is the spectral index of the magnetic turbulence) between the parallel mean free path (MFP) and the rigidity $\rho$ of CRs, which formulates a close link between CR transport behavior and magnetic turbulence properties (\citealt{Hussein2015JGRA}). However, when confronted with a realistic astrophysical environment with strong nonlinear turbulence, the QLT is limited in application. The most obvious drawback of the QLT is that it cannot face the $90^{\circ}$ problem, resulting in theoretically infinite MFPs. To improve the theoretical description of the CR transport in realistic turbulent magnetic fields, various nonlinear theories (NLTs) and the extended QLT have been formulated (e.g., \citealt{Voelk1975RvGSP, Matthaeus2003ApJ, Shalchi2009ASSL, Yan2008ApJ, Xu2018}). In the framework of the NLT, \cite{Qin2002PhDT} numerically proposed the relation of $\lambda_{\parallel} \propto \rho^{0.6}$, which is different from $\lambda_{\parallel} \propto \rho^{1/3}$ of the QLT prediction for Kolmogorov spectrum.

Given that CR diffusion depends on turbulent magnetic fields with anisotropy, it is reasonable to consider the perpendicular and parallel transport of CRs with regard to the (local) mean magnetic field. For the perpendicular diffusion measured with respect to the local mean magnetic field, it has been demonstrated that CRs undergo subdiffusion and superdiffusion in the dissipation range and inertial range, respectively (\citealt{Hu2022MNRAS, Zhang2023ApJ}). In the inertial range, the separation of particles increases with the time as $t^{3/2}$ due to the perpendicular superdiffusion of magnetic field lines (\citealt{Lazarian2014ApJ, Hu2022MNRAS, Sampson2023MNRAS}). When considering the CR parallel diffusion, gyroresonance scattering, describing the interaction of CRs with MHD turbulence (\citealt{Yan2008ApJ, Xu2018}), leads to scattering diffusion in the direction parallel to the local magnetic field. It has been claimed that the fast mode is a dominant scattering agent, and slow and Alfv\'en modes with scale-dependent anisotropy can suppress the CR scattering (\citealt{Yan2002PhRvL}).

Except for the pitch-angle scattering by gyroresonant interactions, CRs are also reflected by magnetic mirrors (\citealt{Fermi1949PhRv, Cesarsky1973ApJ}). Recently, \cite{Xu2020ApJ} analytically predicted the mirroring and scattering effect of CRs in MHD turbulence. In addition, they found that isotropic fast mode dominates both mirroring and gyroresonant scattering of CRs in the case of compressible MHD turbulence, and the mirroring by slow mode dominates over the gyroresonant scattering by Alfv\'en and slow modes in the incompressible case. Given the intrinsic superdiffusion of the turbulent magnetic fields, which has been predicted by \cite{LV99}, and numerically confirmed by \cite{Beresnyak2013ApJ} with a high-resolution simulation (see also \citealt{Lazarian2004ApJ}), \cite{LazarianXu2021} proposed that CRs cannot permanently be trapped within one magnetic mirror but move from one magnetic mirror to another along the magnetic field line, which is called the ``mirror diffusion'' that is slower than the scattering diffusion. In addition, the most obvious advantage of mirror diffusion is that it can solve the $90^{\circ}$ scattering problem. 

Numerically, it has been demonstrated that CRs experience mirror diffusion at relatively large pitch angles and gyroresonant scattering at smaller pitch angles, the pitch-angle-dependent parallel MFP due to mirror diffusion alone being much smaller than the driving scale of MHD turbulence (\citealt{Zhang2023ApJ}). At the same time, it has been found that lower-energy CRs preferentially undergo the mirror and wandering diffusion in the strong-field regions amplified by the nonlinear turbulent dynamo (\citealt{ChaoXu2024ApJ}). Based on MHD simulations of star-forming regions, \cite{Barreto-Mota2024arXiv} also demonstrated that the parallel mirror diffusion dominates the CR diffusion behavior.

With motivations for both slow diffusion found in observations and theoretical predictions of turbulent mirror diffusion, we further numerically investigate mirror and scattering diffusion behavior. One purpose of our work is to investigate potential quantitative relationships between the CR MFPs and their energies. Another purpose is to study how individual MHD modes affect the mirror and scattering diffusion of CRs. The paper is organized as follows. Brief theoretical descriptions of scattering and mirroring are provided in Sect. \ref{sec: style2}. Section \ref{sec: style3} describes our numerical methods, including generating 3D data of MHD turbulence, MHD mode decomposition, and the test-particle method. Numerical results are presented in Sect. \ref{sec: style4}, followed by discussion and summary in Sects. \ref{sec: style5} and \ref{sec: style6}, respectively.

\section{Theoretical description}\label{sec: style2}
MHD turbulence can be described by the ideal MHD equations (see also Sect. \ref{sec: style3}). In general, MHD turbulence can be decomposed into three modes: Alfv\'en, slow, and fast modes (e.g., \citealt{Beresnyak2019LRCA}), numerically achieved by the Fourier (CL02) and wavelet (\citealt{Kowal2010}, hereafter KL10) transformations. It has been numerically demonstrated that Alfv\'en and slow modes have scale-dependent anisotropy, and the fast mode presents isotropy (CL02 and CL03). In addition, Alfv\'en mode is incompressible while fast and slow modes are compressible. 

The interactions of CR particles with turbulent magnetic fields affect the transport of CR particles in the directions parallel and perpendicular to the magnetic field. The resonant interaction of gyroresonance leads to scattering diffusion, while the nonresonant interaction with turbulent magnetic mirrors leads to mirror diffusion. They lead to parallel diffusion together. According to the QLT theory, the linear resonance function of gyroresonance scattering is given by (\citealt{Kulsrud1969ApJ, Voelk1975RvGSP})
\begin{equation}
\begin{aligned}
R_n = \pi \delta(k_\mathrm{\parallel} v_\mathrm{\parallel} - \omega \pm n\Omega ),  \label{eq: Rn}
\end{aligned}
\end{equation}
where $\omega$ and $\Omega$ are the wave frequency and particle gyrofrequency, respectively. This gyroresonance scattering requires that $\omega$ is Doppler-shifted to the gyrofrequency $\Omega$ of the particle or its cyclotron harmonics $n\Omega$ with a nonzero integer $n$ (\citealt{Yan2008ApJ, Xu2018}). When the parallel velocity $v_\parallel$ of particles matches with the wave phase speed $\omega/k_{\parallel}$, the resonance condition corresponds to a simplified resonance function
\begin{equation}
\begin{aligned}
R_n = \pi \delta(k_\mathrm{\parallel} v_\mathrm{\parallel} - \omega),  \label{eq: Rn1}
\end{aligned}
\end{equation}
resulting in efficient particle acceleration, i.e., the so-called transit time damping (TTD) process (\citealt{Yan2002PhRvL, Schlickeiser2002}). It is worth noting that TTD is one of the mechanisms for particle acceleration (\citealt{Yan2015ASSL}). 

In the case of nonresonant interaction, CRs can interact with the multi-scale mirrors naturally generated due to the turbulent compressions of magnetic fields. CR particles undergoing the mirror diffusion satisfy the conditions of $R_\mathrm{g} < l_\mathrm{mir}$ and $\mu< \mu_\mathrm{c}$ (\citealt{LazarianXu2021}), where $R_\mathrm{g}$ is the Larmor radius, $\mu$ the cosine of the pitch angle, and $l_\mathrm{mir} $ the scale of the mirror. The pitch angle is the angle between the velocity of the particle and the magnetic field. Note that the critical value $\mu_\mathrm{c}$ can be obtained by $\mu_\mathrm{c}=\rm{min[\mu_\mathrm{mir},\mu_\mathrm{eq}]}$, where $\mu_\mathrm{mir}\approx \sqrt{\frac{\delta B}{B_0 + \delta B}} \approx \sqrt{\frac{\delta B}{B_0}} $ corresponds to the loss cone angle ($B_0$ and $\delta B$ denotes the mean and fluctuating magnetic fields, respectively), and $\mu_\mathrm{eq} \simeq [\frac{14}{\pi}(\frac{\delta B_\mathrm{f}}{B_0})^2(\frac{R_\mathrm{g}}{L_{\rm inj}})^{\frac{1}{2}}]^{\frac{2}{11}}$ corresponds to $\mu$ value when the mirroring rate equals to scattering rate of particles. In the case of compressible MHD turbulence, due to the fast mode dominating the mirror diffusion (i.e., the mirroring rate of CRs by fast mode greater than that by slow mode), $\mu_\mathrm{mir}$ can be replaced by (\citealt{LazarianXu2021})
\begin{equation}
\begin{aligned}
\mu_\mathrm{mir} = \sqrt{\frac{\delta B_\mathrm{f}}{B_0 + \delta B_\mathrm{f}}}.\label{eq: mu_c}
\end{aligned}
\end{equation}
The mirroring particles preserve the first adiabatic invariant, that is, the conservation of the magnetic moment $M=\frac{\gamma m u_\perp^2}{2B}=\rm{const}$, where $\gamma$, $m$ and $u_\perp$ denote the Lorentz factor, mass and perpendicular velocity of CR particles, respectively.

The perpendicular transport of CRs is influenced by the superdiffusion of the magnetic field. Since perpendicular diffusion of CR depends on the transport scale, one can divide the perpendicular transport process into two parts: the large-scale transport (the scale larger than the injection scale $L_\mathrm{inj}$), and the small-scale one (the scale smaller than $L_\mathrm{inj}$). According to the $M_{\rm A}$ values (see Sect. \ref{sec: style3} for its definition), the turbulence can be divided into super-Alfv\'enic ($M_{\rm A}>1$) and sub-Alfv\'enic ($M_{\rm A}<1$) turbulence regimes. For the former, the transition from hydrodynamic turbulence to MHD turbulence happens at $L_\mathrm{A}=L_\mathrm{inj}M_\mathrm{A}^{-3}$, while for the latter, the transition from weak turbulence to strong turbulence happens at $L_\mathrm{tr}=L_\mathrm{inj}M_\mathrm{A}^{2}$ (see \citealt{BL2019.book} for more details).

The propagation of CRs is related to the magnetization degree of turbulence. In the case of super-Alf\'enic turbulence, on a scale larger than $L_\mathrm{A}$, with turbulent energy dominating the magnetic energy, the diffusion is expected to be isotropic, which means that the parallel diffusion coefficient $D_\mathrm{\parallel}$ and the perpendicular diffusion coefficient $D_\mathrm{\perp}$ are similar (\citealt{Yan2008ApJ, Lazarian2014ApJ})
\begin{equation} 
\begin{aligned}
D_\mathrm{\perp} \approx D_\mathrm{\parallel}. \label{eq: D_large_sup}
\end{aligned}
\end{equation}
However, when the scale is less than $L_\mathrm{A}$, the diffusion is expected to be anisotropic, and the parallel and perpendicular diffusion coefficients follow (\citealt{Maiti2022ApJ})
\begin{equation} 
\begin{aligned}
D_\mathrm{\perp} \approx D_\mathrm{\parallel}M_\mathrm{A}^3.\label{eq: D_less_sup}
\end{aligned}
\end{equation}
For the sub-Alfv\'enic turbulence, the parallel and perpendicular diffusion coefficients maintain the following relation (\citealt{Lazarian2014ApJ})
\begin{equation} 
\begin{aligned}
D_\mathrm{\perp} \approx D_\mathrm{\parallel}M_\mathrm{A}^4,\label{eq: D_large_sub}
\end{aligned}
\end{equation}
which have been numerically confirmed by \cite{Xu2013} and \cite{Maiti2022ApJ}.

\section{Simulation methods} \label{sec: style3}

\begin{table*}
\renewcommand{\arraystretch}{1.5}
\caption{Parameters of MHD turbulence data cubes.}
\centering
\setlength{\tabcolsep}{10mm}
\resizebox{\textwidth}{10mm}{
\begin{tabular}{ccccccccc}
\hline\hline
Models & $M_\mathrm{A}$ & $M_\mathrm{S}$ & $\beta$ & $B_0$  & $\delta{B}_\mathrm {rms}/\langle B \rangle$&$V_\mathrm {rms} /V_\mathrm{A}$ & $\delta{B_\mathrm{f}}/\langle B \rangle$&$\mu_c$\\
\hline
R1  & 1.69   & 3.11   & 0.591  &0.1 & 1.08 &1.81 &0.493 &0.51 \\
R2  & 0.65   & 0.48   & 3.668  &1.0 & 0.54 &0.75 &0.427 &0.49 \\
R3  & 0.58   & 3.17   & 0.067  &1.0 & 0.46 &0.63 &0.496 &0.51\\
\hline\hline
\end{tabular}}
\tablefoot{The physical meaning of symbols are explained as follows: $M_\mathrm{A}$ is the Alfv\'enic Mach number; $M_\mathrm{S}$ sonic Mach number; $\beta$ the plasma parameter; $B_0$ initial magnetic field strength; $\langle B \rangle$ mean magnetic field; $\delta{B}_{\rm rms}$ root mean square of the random magnetic field; $V_\mathrm {rms}$ root mean square of the fluid velocity; $V_\mathrm{A}$ Alfv\'en speed; $\mu_c$ the critical value of the cosine of the pitch angle; and $\delta{B_{\rm f}}$ fluctuation magnetic field of the fast mode.
}
\label{table_data}
\end{table*}

\begin{figure*}[h]
	\begin{minipage}{0.5\linewidth}
		\vspace{3pt}
		\centerline{\includegraphics[width=\textwidth]{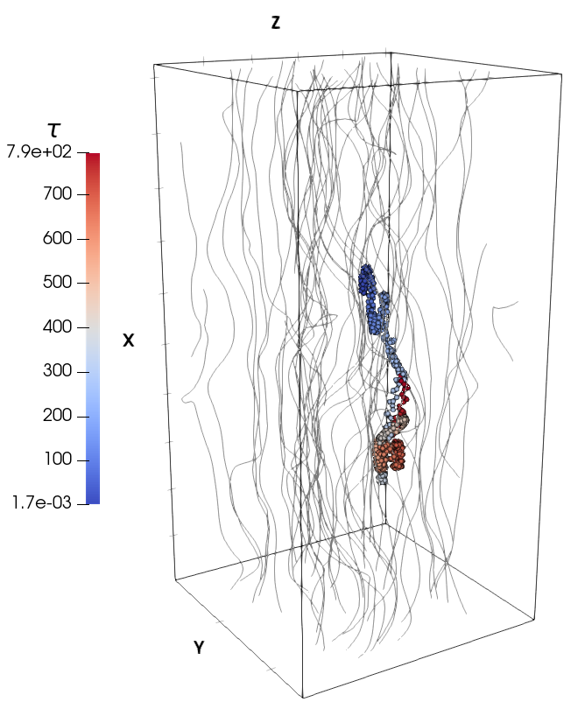}}
	\end{minipage}
	\begin{minipage}{0.43\linewidth}
		\vspace{3pt}
		\centerline{\includegraphics[width=\textwidth]{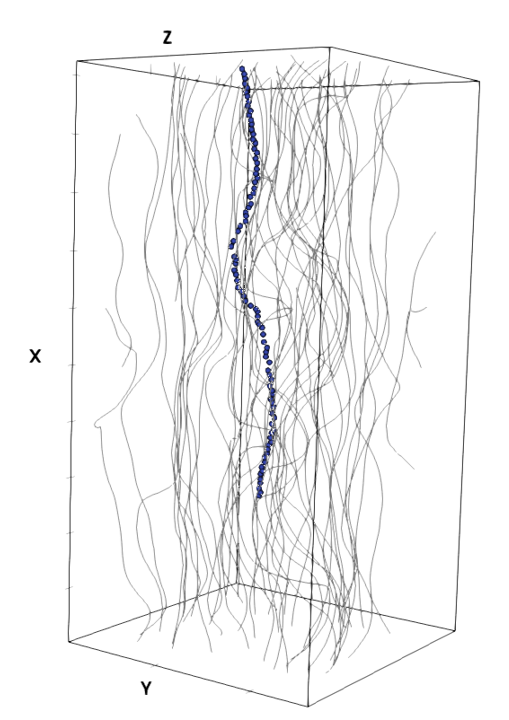}}
	\end{minipage}
	\caption{The trajectories of the initially mirroring particle (left panel; initial pitch-angle cosine $\mu_0=0.15$) and the initially scattering particle (right panel; $\mu_0=0.8$). The magnetic field lines (black) extend to 1024 pixels in the $x$-axis direction. The trajectories of these two particles are color-coded by the time $\tau$ (normalized by the gyrofrequency $\Omega$), as shown in the color bar. Note that for the sake of comparison, the right panel only shows a part of the trajectory in the range of $\tau <160$. The Larmor radius of these two particles is $R_\mathrm{g} = 0.03L_\mathrm{inj}$. Simulations are from R2 listed in Table \ref{table_data}. }
	\label{fig: particle_traj_3D}
\end{figure*}
 
\begin{figure}
	\centering  
	\subfigure{
		\label{level.sub.1}
		\includegraphics[width=1\linewidth]{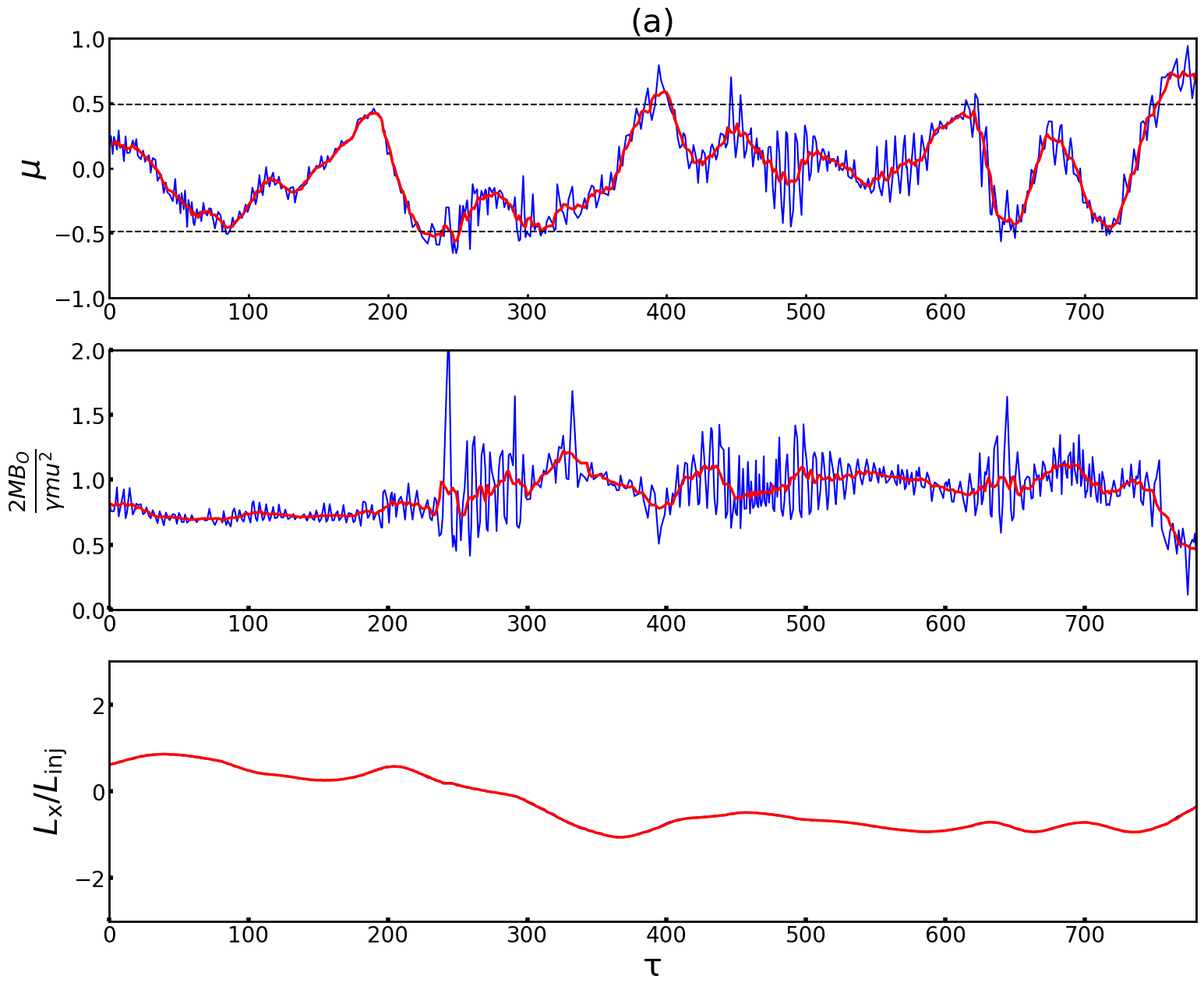}}
        \quad
	\subfigure{
		\label{level.sub.2}
		\includegraphics[width=1\linewidth]{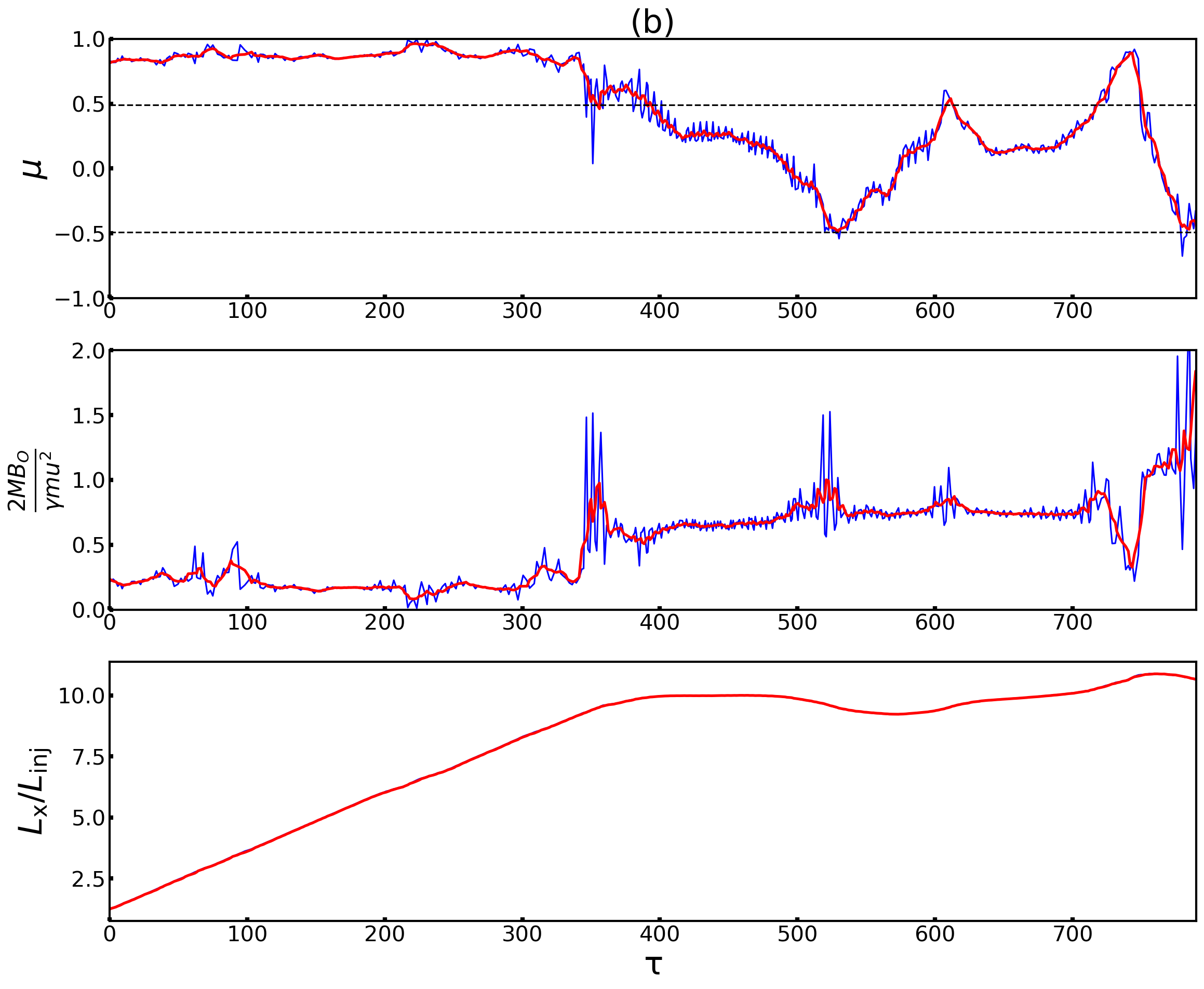}}
\caption{The cosine of pitch angle $\mu$, normalized magnetic moment $2MB_0/\gamma m u^2$, and spatial displacement $L_\mathrm{x}/L_\mathrm{inj}$ in the $x$-axis direction as a function of time $\tau$. The results of initially mirroring and scattering particles shown in panels (a) and (b) correspond to the left and right particles of Fig. \ref{fig: particle_traj_3D}, respectively. The horizontal dashed lines represent $\pm \mu_\mathrm{c}$ in Table \ref{table_data}. The blue and red curves represent the numerical result and the averaged one within one gyration period, respectively.}
	\label{fig: M_mu_t}
\end{figure}

To model turbulent magnetic fields, we perform 3D numerical simulations to solve the following ideal MHD equations:
\begin{equation}
\frac{\partial\rho}{\partial t} +\nabla\cdot (\rho
\bm{v}) = 0,\label{eq: rho}
\end{equation}
\begin{equation}
\rho\left[\frac{\partial \bm{v}}{\partial t} + (\bm{v} \cdot \nabla)\bm{v}\right] + \nabla p - \frac{1}{4\pi} \bm{J} \times \bm{B} = \bm{f},\label{eq: mom}
\end{equation}
\begin{equation}
\frac{\partial \bm{B}}{\partial t} - \nabla \times (\bm{v} \times \bm{B}) = 0,\label{eq: mag}
\end{equation}

\begin{equation}
\nabla \cdot \bm{B} = 0, \label{eq: magnetic_gauss}
\end{equation}
where $\bm{f}$ is a random force to drive the turbulence, $\bm{J} = \nabla \times \bm{B}$ the current density, and $\bm {v}$ the fluid velocity. In Eq. (\ref{eq: mom}), $p=c_\mathrm{s}^2\rho$ represents gas pressure, where $c_\mathrm{s}$ and $\rho$ are the sonic speed and density, respectively.

We use a second-order-accurate hybrid simulation code developed by \cite{CL02}, which is essentially non-oscillatory. Considering a periodic boundary condition and setting a non-zero mean magnetic field in the $x$-axis direction, we drive 3D simulations by a pure solenoidal forcing in the Fourier space. With the numerical resolution of $L^3=512^3$, we set the injection wave number $k_{\rm in}\simeq 2.5$, corresponding to the injection scale of $0.4L$. When the simulation reaches a statistically steady state (with a steady turbulent energy spectrum), we take 3D data cubes with density, velocity, and magnetic field. Defining the Alfv\'enic Mach number $M_\mathrm{A} = \left\langle\frac{| \bm{v}|}{V_\mathrm{A}}\right\rangle$ and the sonic Mach number $M_\mathrm{s} = \left\langle\frac{| \bm{v}|}{c_\mathrm{s}}\right\rangle$, we can obtain the values of characteristic parameters (listed in Table \ref{table_data}) of each simulation. Here, the Alfv\'en speed $V_\mathrm{A}$ is related to the total magnetic field strength and plasma density by $V_\mathrm{A} = \frac{|B|}{\sqrt{\rho}}$. In addition, the plasma parameter $\beta = p_{\rm gas}/p_{\rm mag} = 2M_{\rm A}^2/M_{\rm s}^2$ is defined as the ratio of gas pressure to magnetic pressure.

We decompose the compressible MHD turbulence into Alfv\'en, slow, and fast modes by the wavelet transformations (KL10). The displacement vectors of three modes are defined by (CL02; KL10)
\begin{equation}
\hat{\xi_\mathrm{f}} \propto (1 + \alpha + \sqrt{D})k_{\perp}\hat{\bm{k}_{\perp}} + (-1 + \alpha + \sqrt{D})k_{\parallel}\hat{\bm{k}_{\parallel}},\label{eq: fast}
\end{equation}
\begin{equation}
\hat{\xi_\mathrm{s}} \propto (1 + \alpha - \sqrt{D})k_{\perp}\hat{\bm{k}_{\perp}} + (
-1 + \alpha - \sqrt{D})k_{\parallel}\hat{\bm{k}_{\parallel}},\label{eq: slow}
\end{equation}
\begin{equation}
\hat{\xi_\mathrm{A}} \propto \hat{\bm{k}_{\perp}} \times \hat{\bm{k}_{\parallel}},\label{eq: Alf}
\end{equation}
where $D = (1 + \alpha )^2 - 4\alpha\cos^2{\theta}, \alpha = c_\mathrm{s}^2/V_{\rm A}^2$ and $\theta$ is the angle between the wave vector $\bm{k}$ and $\bm{B_0}$. We first use the discrete wavelet transform to obtain the wavelet coefficients corresponding to the magnetic field. Next, we perform the Fourier decomposition for three modes following CL02 by projecting the magnetic field into the displacement vectors of three modes (see Eqs. (\ref{eq: fast}) to (\ref{eq: Alf})) in the Fourier space. Finally, we carry out the inverse Fourier transform for three modes to get the information on the magnetic field in real space.

By injecting test particles into 3D data cubes, we can simulate the transport of CR particles. Numerically, we use the Bulirsch–Stoer method \citep{Press1986book} with an adaptive time step to trace the motions of the charged particles. Specifically, the motions of test particles are governed by the Lorentz equation as follows
\begin{equation}
\frac{d}{dt}(\frac{m {\bm u}}{\sqrt{(1-u^2/c^2)}} ) = q( {\bm u} \times {\bm B}),\label{eq: Lor_force}
\end{equation}
where $q$ and $\gamma=\frac{1 }{\sqrt{(1-u^2/c^2)}}$ represent the charge and Lorentz factor of the particle, respectively. The particle position $\bm r$ is related to its velocity by 
\begin{equation}
\frac{d\bm r}{dt}=\bm u, \label{eq: position}
\end{equation}
where the time $t$ in code units is normalized as $\tau = t \cdot \Omega$ and $\Omega = u/R_\mathrm{g}$. In our simulation, we use the ratio of the Larmor radius $R_\mathrm{g}$ to $L_\mathrm{inj}$ to characterize test particle energy, where $R_\mathrm{g}$ is defined as
\begin{equation}
R_\mathrm{g}=\gamma \frac{m c^2}{q B_0}.\label{eq: rg}
\end{equation}

To numerically describe the diffusion processes of CR particles, we define the parallel diffusion coefficient $D_\mathrm{\parallel}$ and the perpendicular one $D_\mathrm{\perp}$ with respect to the mean magnetic field direction as 
(\citealt{Giacalone1999ApJ})
\begin{equation}
\begin{aligned}
D_\mathrm{\parallel} = \frac{\langle \Delta \ell_\mathrm{\parallel}^2 \rangle}{2 \Delta t} \label{eq: D_p}
\end{aligned}
\end{equation}
and
\begin{equation} 
\begin{aligned}
D_\mathrm{\perp} = \frac{\langle \Delta \ell_\mathrm{\perp}^2 \rangle}{2 \Delta t}, \label{eq: D_v}
\end{aligned}
\end{equation}
respectively. In Eqs. (\ref{eq: D_p}) and (\ref{eq: D_v}), the particle's parallel and perpendicular displacements squared with respect to the mean magnetic field are defined as  
\begin{equation}
\Delta \ell_\mathrm{\parallel}^2= (x-x_0)^2 \label{eq: dx_p}
\end{equation}
and
\begin{equation}
\Delta \ell_\mathrm{\perp}^2= (y-y_0)^2+(z-z_0)^2 , \label{eq: dx_v}
\end{equation}
respectively. The symbol $\langle \dots \rangle$ denotes the average over all particles. Furthermore, we can define the parallel and perpendicular MFPs of particles as 
\begin{equation}
\lambda_\parallel = \frac{3 D_\parallel}{u}\  \ {\rm and }\  \ \lambda_\perp = \frac{3 D_\perp}{u}, \label{eq_mfps}
\end{equation}
respectively.

\section{Numerical results} \label{sec: style4}
\subsection{Spatial diffusion of single particle}
Before exploring the statistical properties of particle diffusion, we here focus on the behaviors of the individual initially mirroring and scattering particles. Figure \ref{fig: particle_traj_3D} depicts the trajectories of the randomly selected initially mirroring (left panel) and initially scattering (right panel) particles. We would like to stress that the mirroring and scattering particles are prepared initially, but at later times, they will lose their identification as $\mu$ changes through scattering. Due to the setting of the mean magnetic field along the $x$-axis direction, we see that the curved magnetic field lines (thin solid lines) distribute along the $x$ axis. We set the initial pitch-angle cosines $\mu_0=0.15$ and $\mu_0=0.8$ corresponding to the initially mirroring and scattering particles, respectively. Their trajectories are color-coded by time $\tau$. Compared to these two particles in Fig. \ref{fig: particle_traj_3D}, we can see that the parallel spatial extent of the initially mirroring particle (left panel) is smaller than that of the initially scattering particle (right panel). Physically, the mirroring effect can more effectively confine particles compared with scattering. Due to the perpendicular superdiffusion of magnetic field lines, mirroring particles can interact with different magnetic mirrors by bouncing back and forth among several different magnetic mirrors. At a later time, the initially mirroring particle changes its pitch angle due to the scattering. When the pitch angle decreases to a sufficiently small value, i.e., close to the condition of $\mu >\mu_\mathrm{c}$, the mirroring particle escapes from turbulent magnetic mirrors and undergoes scattering diffusion. When the pitch angle becomes sufficiently large with $\mu < \mu_\mathrm{c}$, it will again undergo the mirror diffusion.

To observe the detailed feature of these two particles, we plot $\mu$, magnetic moment $M$, and the position in the $x$-axis direction $L_\mathrm{x}$ as a function of the time in Fig. \ref{fig: M_mu_t}. Panels (a) and (b) correspond to the initially mirroring and scattering particles in Fig. \ref{fig: particle_traj_3D}, respectively. As seen in panel (a), the initially mirroring particle undergoes mirror diffusion before $\tau \simeq 220$ with an insignificant change in $L_\mathrm{x}$ and the pitch angle crossing 90 degrees over and over again. When $\mu$ approaches the critical condition $\mu_c$, the particle suffers from a weak scattering process, making it escape from the current mirror and enter another mirror. Again, the particle experiences a similar mirror diffusion in the range from $\tau \simeq 300$ to 700. We note that during the mirror diffusion process, the (normalized) magnetic moment maintains $M\simeq 1.0$. After $\tau \simeq 750$, the particle experiences a significant scattering, which allows the particle to leave the mirrors and then undergoes scattering diffusion. If one traces the particle long enough, it can undergo mirror diffusion again when the conditions of mirror diffusion: $\mu <\mu_c$ and $M=\rm const$ are satisfied. It is found that the combination of the mirror and scattering diffusion results in a L\'evy-flight-like propagation for CRs (see also \citealt{Barreto-Mota2024arXiv} for more details).

As seen in panel (b), the initially scattering particle rapidly diffuses a long distance along the magnetic field before $\tau \simeq 300$ while $\mu$ and $M$ almost maintain their initial values, i.e., $\mu \simeq 0.8$ and $M \simeq 0.3$, respectively. This indicates inefficient scattering, and thus the particle simply travels along field lines. It is worth noting that for the initially mirroring and scattering particles, they both remain constant in the magnetic moment at the early stage of time, but $M\simeq 0.3$ for the initially scattering particle is different from $M\simeq 1.0$ for the initially mirroring particle. Therefore, one cannot only use the constant magnetic moment to identify a mirroring particle\footnote{When a particle diffusion in MHD turbulence with a constant $M$ and a large $\mu$, we usually call it wandering.}. In the range from $\tau \simeq 400$ to 700, the particle satisfies the conditions of mirror diffusion mentioned above, with the insignificant change of $L_\mathrm{x}$. After $\tau \simeq 700$, the particle shortly undergoes the scattering diffusion and then back to the mirror diffusion region. 

\subsection{Parallel and perpendicular displacements of particles}
\begin{figure*}
   \centering
   \includegraphics[width=\hsize]{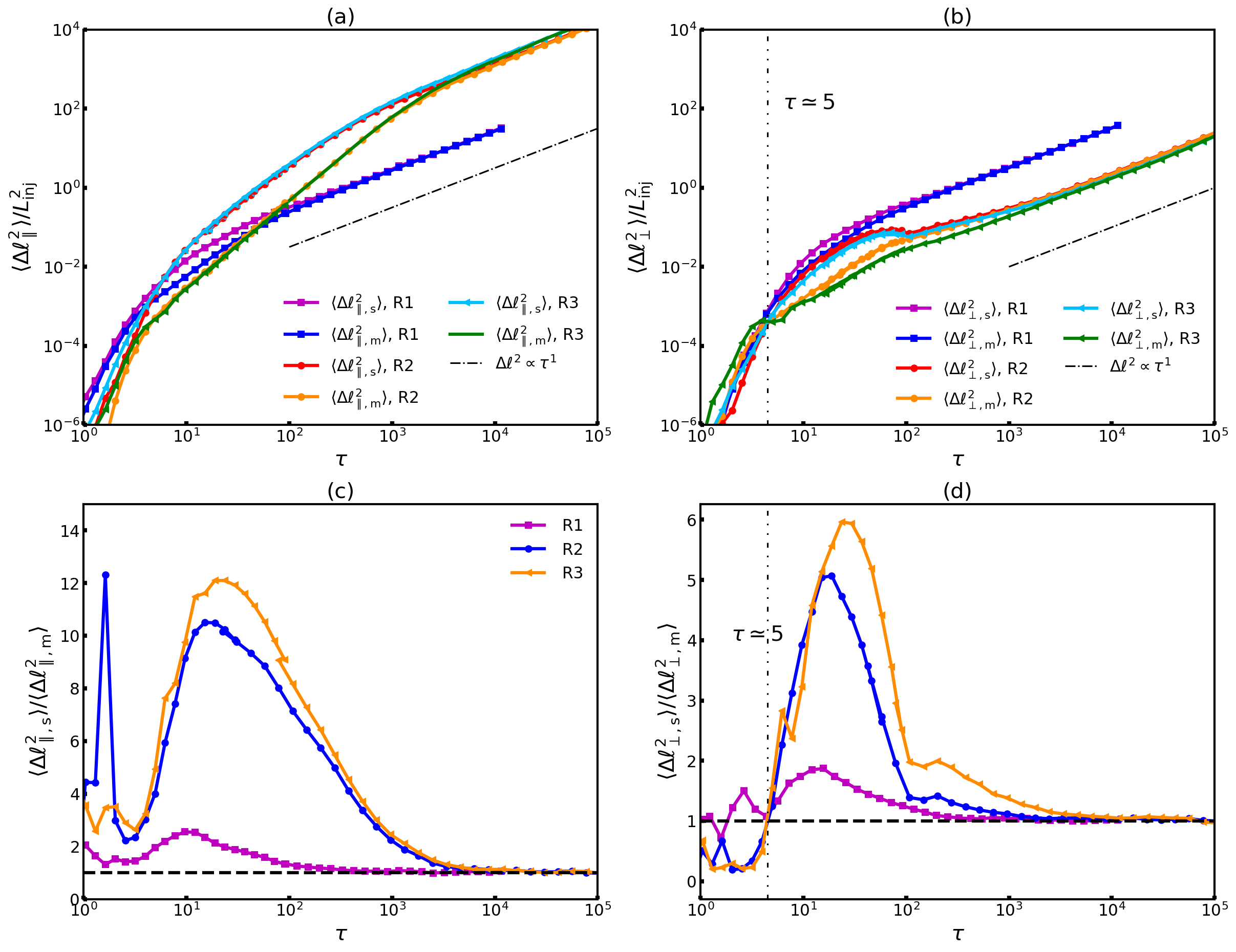}
      \caption{Upper panels: the mean square displacement traveled by overall initially mirroring and scattering particles versus the time (normalized by the gyrofrequency $\Omega$) in the parallel (panel a) and perpendicular (panel b) directions with respect to the mean magnetic fields. Lower panels: the ratio of the parallel (panel c) and perpendicular (panel d) mean square displacement between the initially scattering and mirroring particles. The initial pitch angle values are set to $\mu_0 \in [0.1,0.2]$ and $\mu_0 \in [0.8,0.9]$ for initially mirroring and scattering particles, respectively. The Larmor radius of particles is $R_\mathrm{g} = 0.03L_{\rm inj}$. }
         \label{fig: particle_dx}
   \end{figure*}

This section explores statistically the diffusion properties of a large sample of initially mirroring and scattering particles. Figure \ref{fig: particle_dx} shows the results of the parallel displacement $\langle \Delta \ell_\mathrm{\parallel}^2 \rangle$ (panel a) and the perpendicular one $\langle \Delta \ell_\mathrm{\perp}^2 \rangle$ (panel b) of the initially mirroring and scattering particles as a function of time and the ratio of $\langle \Delta \ell_\mathrm{\parallel}^2 \rangle$ and $\langle \Delta \ell_\mathrm{\perp}^2 \rangle$ (panels c and d) for them, using the models on R1, R2 and R3 listed in Table \ref{table_data}. We set $\mu_0 \in [0.1,0.2]$ and $\mu_0 \in [0.8,0.9]$ for the initially mirroring and scattering particles with the same Larmor radius $R_\mathrm{g}=0.03 L_\mathrm{inj}$, respectively. In both cases, we inject 2000 test particles to trace their trajectories. After particles undergo thousands of gyroperiods, they reach a normal diffusion regime\footnote{In general, if one particle's the mean square displacement $\langle \Delta \ell^2 \rangle$ satisfies a power law relationship of $\langle \Delta \ell^2 \rangle \propto (\Delta t)^{\alpha}$ over the time, it can be called subdiffusion ($\alpha < 1$), normal diffusion ($\alpha = 1$), or superdiffusion ($\alpha > 1$; \citealt{Shalchi2009ASSL}).}. We terminate our test particle simulation. 

In the case of parallel diffusion (see panels a and c), the particles' diffusion can be divided into two stages in terms of the dependence of mean square displacement on time. The first stage is superdiffusion with a power-law index larger than 1. The second one is normal diffusion, where the mean square displacement presents a linear dependence on time of $\langle \Delta \ell_\parallel ^2\rangle \propto \tau$. In the first stage, we can see $\langle \Delta \ell_{\rm \parallel, m}^2 \rangle < \langle \Delta \ell_{\rm \parallel, s}^2 \rangle $ for three cases from panel (c), which means the diffusive displacement of the initially mirroring particles is smaller than that of the initially scattering particles during the same time. This is because the initially mirroring particles are more effectively confined, compared to initially scattering particles. It means that the presence of magnetic mirrors makes the mirror diffusion confine CRs more strongly than scattering diffusion. In the second stage, we can see $\langle \Delta \ell_{\rm \parallel, m}^2 \rangle \simeq \langle \Delta \ell_{\rm \parallel, s}^2 \rangle $. It indicates that their displacements converge, irrespective of the initial pitch angles, when the parallel diffusion undergoing normal diffusion. Compared with different turbulence regimes, the main factor affecting parallel diffusion is $M_{\rm A}$ rather than $M_{\rm s}$. As seen in the first stage, the initially mirroring and scattering particles have a larger number of gyrations and faster parallel diffusion for sub-Alfv\'enic turbulence (see R2 and R3 listed in Table \ref{table_data}) than that for super-Alfv\'enic turbulence (see R1). As for the case of R1, due to the enhanced fluctuation of the magnetic field, particles are subjected to more efficient scattering of the magnetic field, the initially mirroring and scattering particles earlier to reach normal diffusion.

In the case of perpendicular diffusion (see panels b and d), except for the presence of a significant oscillation at $\tau \lesssim 5$, which may be from numerical dissipation, we can also see $\langle \Delta \ell_{\rm \perp,  m}^2 \rangle<\langle \Delta \ell_{\rm \perp, s}^2 \rangle $ for superdiffusion and $\langle \Delta \ell_{\rm \perp, m}^2 \rangle \simeq \langle \Delta \ell_{\rm \perp, s}^2 \rangle $ for normal diffusion in the different turbulence regimes. Compared with R1, R2, and R3 within the same time, the initially mirroring and scattering particles all diffuse larger distances for R1 than those for R2 and R3, for which the distance is comparable. In addition, we note that R1 shows the isotropic diffusion of $\langle \Delta \ell_{\rm \perp}^2 \rangle \approx \langle \Delta \ell_{\rm \parallel}^2 \rangle $ during the whole process, whereas R2 and R3 indicate anisotropic diffusion of $\langle \Delta \ell_{\rm \perp}^2 \rangle < \langle \Delta \ell_{\rm \parallel}^2 \rangle $. The reason is that the perpendicular diffusion of particles is subject to that of magnetic fields. With increasing the fluctuation of the magnetic field, as characterized via $\delta B_{\rm rms}/\langle B \rangle$ listed in Table \ref{table_data}, the diffusion of CRs would exhibit isotropic property, which is consistent with the expected isotropic diffusion of CRs for super-Alfv\'enic turbulence.

\begin{figure*}
   \centering
   \includegraphics[width=\hsize]{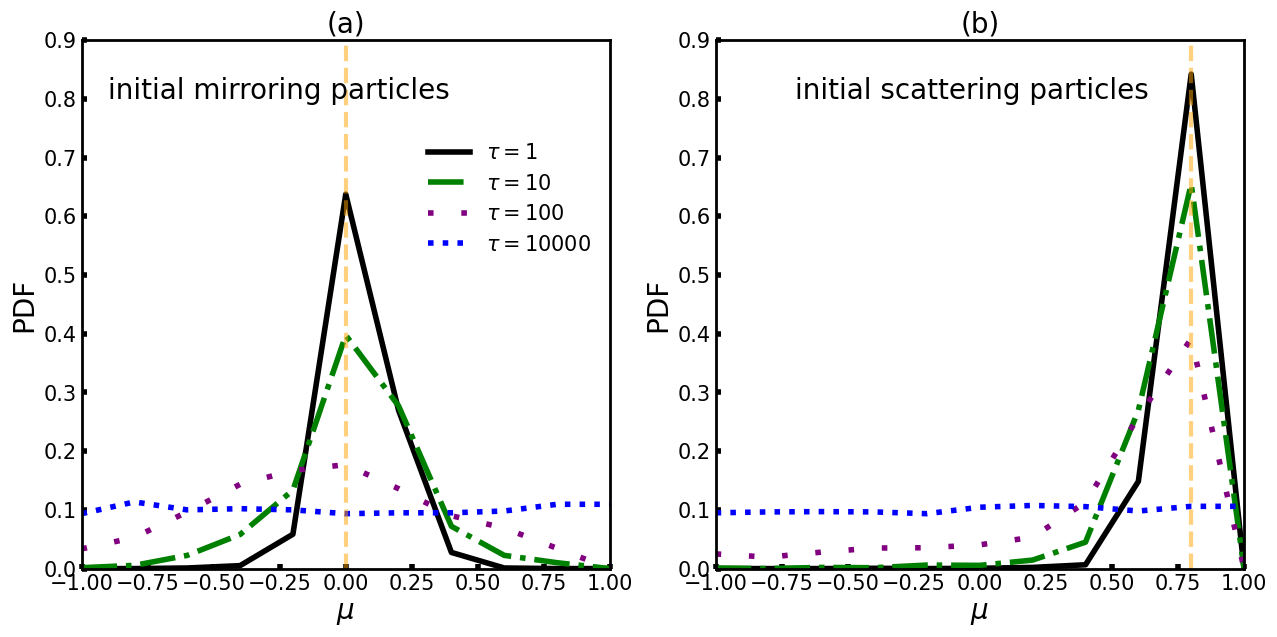}
      \caption{Probability density function of $\mu$ at different times $\tau$, arising from R2 in Fig. \ref{fig: particle_dx}. The vertical dashed lines correspond to $\mu=0$ in panel (a) and $\mu=0.8$ in panel (b).}
         \label{fig: PDF}
   \end{figure*}

To further understand the effect of scattering on the distribution of particle pitch angles, we plot the distribution of $\mu$ corresponding to the initially mirroring particles and scattering particles at different times during the diffusion process. Using model R2 listed in Table \ref{table_data}, we plot the probability density function of $\mu$ in Fig. \ref{fig: PDF} corresponding to Fig. \ref{fig: particle_dx}. As shown, at the first gyro period, i.e., $\tau =1$, the peak of $\mu$ happens at $\mu \simeq 0.0$ for initially mirroring particles (see panel a) and $\mu \simeq 0.8$ for initially scattering particles (see panel b). These two peaks approximately approach the initial $\mu$ distribution that we set. With increasing the time up to $\tau=10^4$, we see that $\mu$ is evenly distributed between $-1$ and $1$ for the initially mirroring and scattering particles, that is, the distribution of $\mu$ is gradually randomized. This demonstrates that, due to scattering, irrespective of the initial pitch angle, particles stochastically undergo mirroring and scattering in turn, resulting in similar diffusion behavior over a sufficiently long time. 

\subsection{Mean free path of particles }

\begin{figure*}
\includegraphics[width=1\linewidth]{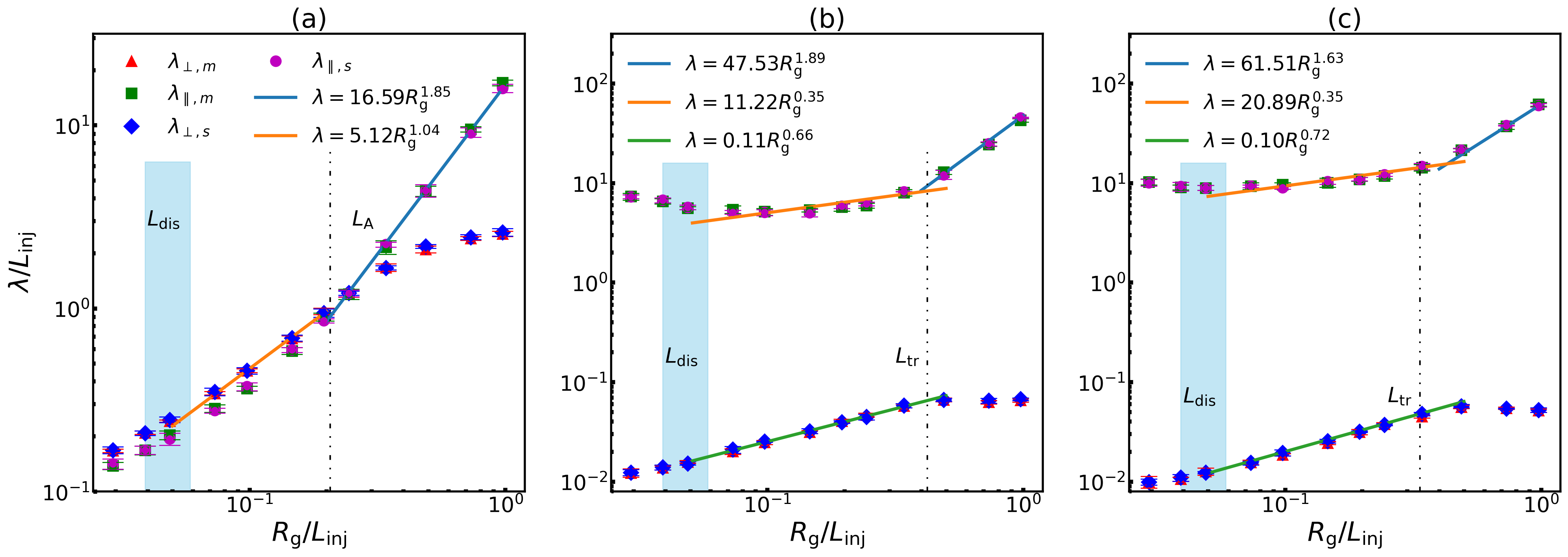}
\caption{The MFPs of mirroring and scattering particles measured at different CR energies. The initial pitch angle values are set to $\mu_0 \in [0.1,0.2]$ and $\mu_0 \in [0.8,0.9]$ for mirroring and scattering particles, respectively. The results in panels (a), (b), and (c) are based on the R1, R2, and R3 listed in Table \ref{table_data}, respectively. The different symbols denote the numerical results, and the solid lines are a linear fitting for them. The vertical dash-dotted lines indicate the transition scales. The blue area plotted in each panel represents the dissipation scales estimated by the power spectra of the magnetic field.
\label{fig: var_rg_MFP}}
\end{figure*}

In this section, we investigate the dependence of the propagation of initially mirroring and scattering particles on their energies in different turbulence regimes. Figure \ref{fig: var_rg_MFP} plots the parallel and perpendicular MFPs, $\lambda_\parallel$ and $\lambda_\perp$ (see Eq. (\ref{eq_mfps})), as a function of the Larmor radius $R_{\rm g}$, where the error bars plotted represent the standard deviation of the MFPs. With the same initial setup of $\mu$ as Fig. \ref{fig: particle_dx}, we run a series of simulations by changing $R_{\rm g}$ values from the dissipation scale (several grids) to the injection scale $L_\mathrm{inj}$. In the early stage (around several hundred gyroperiods), we see superdiffusion for all simulations. During the middle and later stages, where the initially mirroring and scattering particles reach the normal diffusion, i.e., $\langle \Delta \ell^2\rangle \propto \tau$, we measure the parallel and perpendicular MFPs via the parallel and perpendicular diffusion coefficient (see also Sect. \ref{sec: style3}).

In the case of the super-Alfv\'enic turbulence, the numerical result is shown in Fig. \ref{fig: var_rg_MFP}a, from which we see that the relation between $\lambda$ and $R_\mathrm{g}$ presents two significant features separated by the scale $L_{\rm A}$ describing the transition from the hydrodynamic turbulence to strong MHD turbulence (see Sect. \ref{sec: style2}). In the range of $ L_\mathrm{dis} < R_\mathrm{g} < L_\mathrm{A}$, the linear fitting for parallel and perpendicular MFPs vs. $R_{\rm g}$ shows the relation of $\lambda_\mathrm{\parallel} \simeq \lambda_\mathrm{\perp } = 5.12 R_\mathrm{g}^{1.04} \propto R_{\rm g} $. Similarly, we can also see that in the dissipation region of $R_\mathrm{g} < L_\mathrm{dis}$, the parallel and perpendicular MFPs are approximately proportional to $R_{\rm g}$. Given that the random magnetic field dominates the regular magnetic field, $\delta B_{\rm rms}/\langle B \rangle \simeq 1.08$, in this simulation, a strong random fluctuation of the magnetic field lines leads to the absence of anisotropic diffusion.

In the high energy range of $R_\mathrm{g} > L_\mathrm{A}$, we observe a significant difference between the parallel and perpendicular MFPs. The former satisfies a steeper power-law relationship of $\lambda_\mathrm{\parallel}= 16.59R_\mathrm{g}^{1.85}\propto R_{\rm g}^2$, while the latter reaches a plateau-like with increasing energy. Here, the turbulence cascade essentially retains its fluid-like properties. The plateau-like distribution of $\lambda_\mathrm{\perp}$ and the steeper power-law relation of $\lambda_\mathrm{\parallel}$ vs. $R_{\rm g}$ may be due to the weak magnetic field having a marginal influence on turbulence. In addition, the plateau-like distribution of $\lambda_\mathrm{\perp}$ vs. $R_{\rm g}$ can also be associated with the magnetic field wandering. Compared with the initially scattering and mirroring particles with the same energy, we note that the parallel or perpendicular MFPs maintain the same. It indicates that the initial $\mu$ of particles does not change the MFPs of particles.

As for the sub-Alfv\'enic turbulence, our results are plotted in Fig. \ref{fig: var_rg_MFP}b and c. As is shown, the perpendicular MFPs present a power-law distribution of $\lambda_\perp \propto R_{\rm g}^{0.7}\propto R_{\rm g}^{2/3}$ in the low energy range, following a plateau beyond the transition scale $L_{\rm tr}$. This power-law relation is slightly shallower than $\lambda_\perp\propto R_{\rm g}$ shown in Fig. \ref{fig: var_rg_MFP}a. Generally, the power-law relation of the perpendicular diffusion maintains similarity between sub-Alfv\'enic and super-Alfv\'enic turbulence. The significant difference is from the parallel diffusion of particles. First, the relation between $\lambda_\parallel$ and $R_\mathrm{g}$ presents an inverse distribution in the range of the dissipation scale, i.e., the negative index of $\lambda_{\rm\parallel}$ vs. $R_{\rm g}$. This may be associated with the propagation of CRs in damped turbulence (e.g., \citealt{Yan2008ApJ}). Second, the linear fitting provides a power-law relation of $\lambda \propto R_{\rm g}^{0.35} \propto R_{\rm g}^{1/3}$ in the range of strong turbulence of $ L_\mathrm{dis} < R_\mathrm{g} < L_\mathrm{tr}$. Third, the diffusion of the initially mirroring and scattering particles exhibits a distinctly anisotropic characteristic, with the parallel MFPs greater by 2-3 orders of magnitude than the perpendicular ones at the same CR energy.

Similar to Fig. \ref{fig: var_rg_MFP}a in the range of $R_\mathrm{g} > L_\mathrm{tr}$, Fig. \ref{fig: var_rg_MFP}b and c present a steep power-law of $\lambda_{\parallel} \propto R_{\rm g}^{\epsilon}$, with $\epsilon\simeq 1.89$ (panel b) and 1.63 (panel c). The plateau distribution of $\lambda_{\perp}$ demonstrates that in the weak turbulence regime, the parallel diffusion of initially scattering and mirroring particles behaves similarly due to the wandering of the field lines. As a result, we find that the perpendicular and parallel diffusion of initially mirroring and scattering particles strongly depend on the magnetized parameter $M_\mathrm{A}$ of MHD turbulence.

\subsection{Influence of MHD modes on spatial diffusion}

\begin{figure*}\includegraphics[width=1\linewidth]{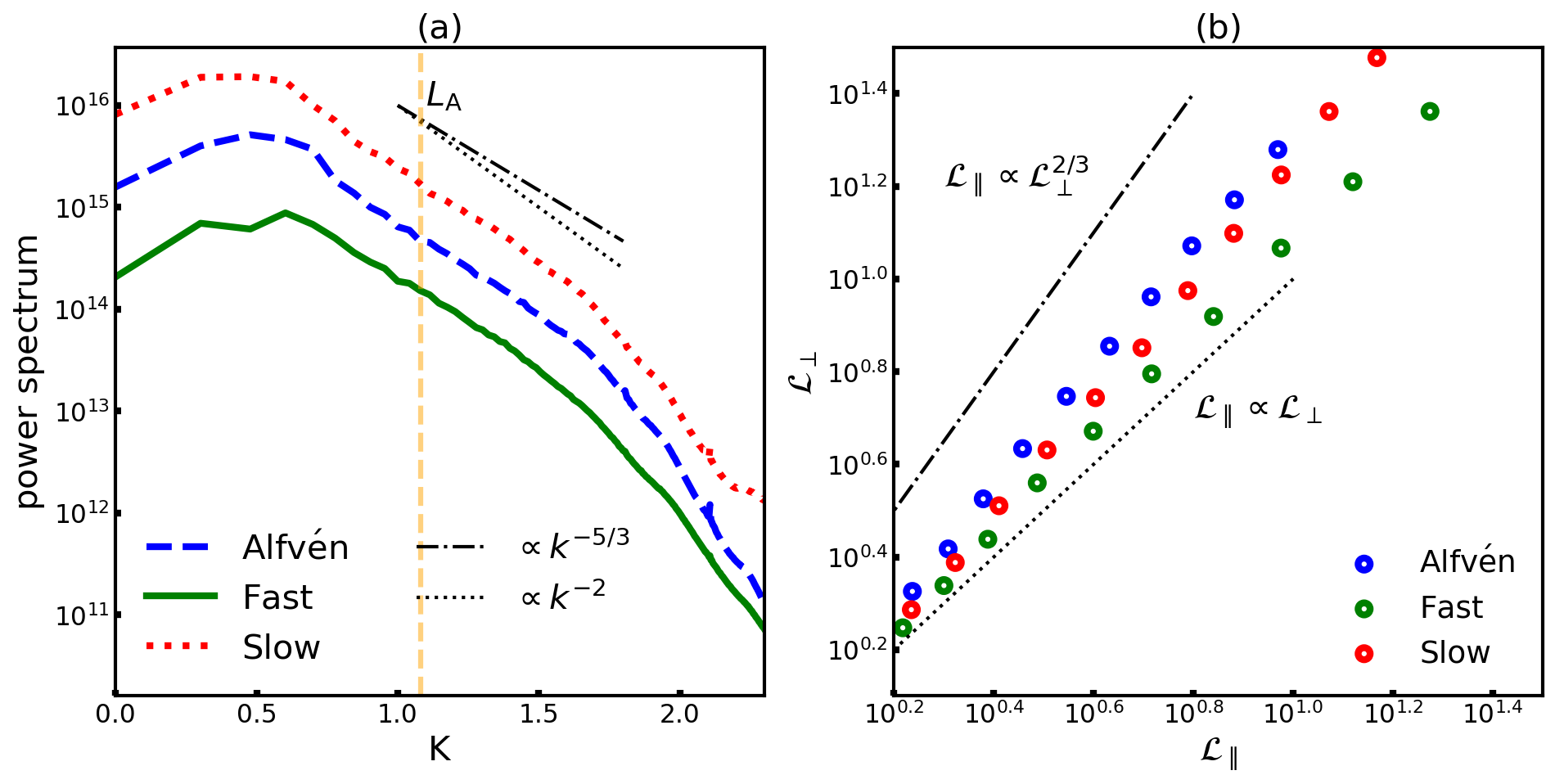}
\caption{The power spectra (panel a) and anisotropy scalings (panel b) of magnetic fields corresponding to the Alfv\'en, fast, and slow modes decomposed from R1.
\label{fig: mode-ps}}
\end{figure*}

\begin{figure*}\includegraphics[width=1\linewidth]{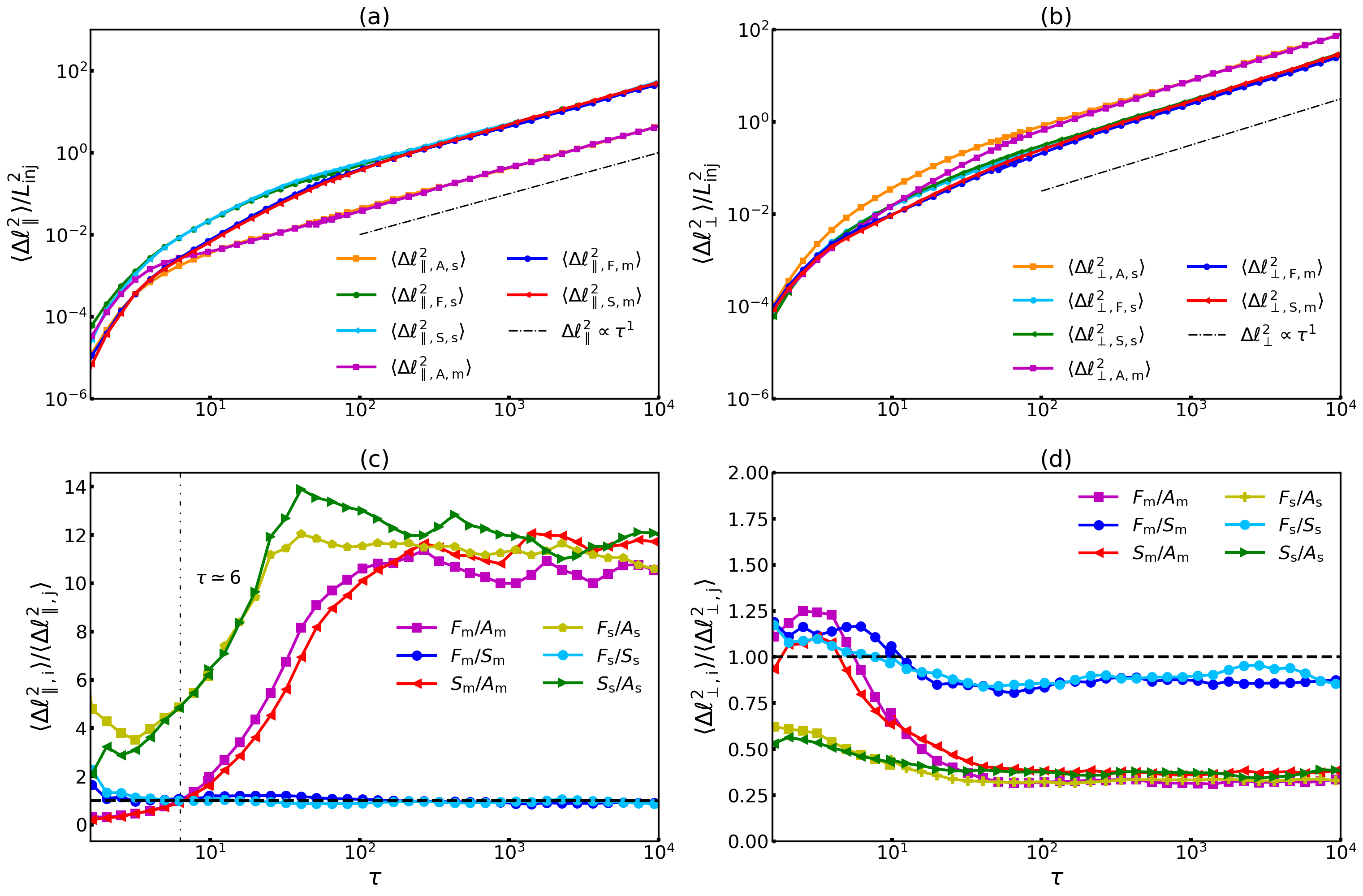}
\caption{Upper panels: the mean square displacement of the parallel (left) and perpendicular (right) diffusion of the initially mirroring and scattering particles versus the time (normalized by $\Omega$) for the Alfv\'en (A), fast (F), and slow (S) modes. Lower panels: the ratio of the parallel (left) and perpendicular (right) mean square displacement between their three plasma modes. The subscripts $i$ and $j$ represent Alfv\'en, fast, and slow modes. Simulations with the Larmor radius of $R_{\rm g} = 0.03L_{\rm inj}$ are based on the decomposed R1 data. The $\mu$ values for initially mirroring and scattering particles are set to $\mu_0 \in [0.1, 0.2]$ and $\mu_0 \in [0.8, 0.9]$, respectively. 
\label{fig: diff_mode}}
\end{figure*}

To explore the influence of different MHD modes on diffusion processes of the initially mirroring and scattering particles, we first decompose 3D MHD turbulence data (R1 as an example) using the wavelet decomposition method (see Sect. \ref{sec: style3}) and then inject test particles into the post-decomposed data of individual modes in the same procedure mentioned above. As shown in Fig. \ref{fig: mode-ps}, the Alfv\'en and slow modes present the anisotropy scaling of $\mathcal{L}_{\parallel}\propto \mathcal{L}_{\perp}^{2/3}$, with the power spectrum of $E\propto k^{-5/3}$. The fast mode has an isotropy scaling of $\mathcal{L}_{\parallel}\propto \mathcal{L}_{\perp}$, with the index of power spectrum close to $-2$ in the inertial range due to the shockwave. In addition, the magnetic energies of three modes satisfy the following relationship $E_{\rm B,S} > E_{\rm B,A} > E_{\rm B,F}$.

Figure \ref{fig: diff_mode}a and b show the evolution of the parallel and perpendicular mean square displacements of the initially mirroring and scattering particles. As shown, the parallel and perpendicular diffusion of particles in three modes experience the transition from superdiffusion to normal diffusion, similar to those in the pre-decomposed case. The magnetosonic modes dominate the parallel diffusion processes for the initially mirroring and scattering particles (see panel a), and the Alfv\'en mode dominates the perpendicular diffusion processes (see panel b). Compared with initially mirroring and scattering particles, we note that significant differences occur in the early stages of time ($\tau<10^2$).

Moreover, in Fig. \ref{fig: diff_mode}c and d, we plot the ratio of the parallel and perpendicular mean square displacements between different modes, respectively. In the case of parallel diffusion (panel c), we can see $\langle \Delta \ell_{\rm \parallel, F}^2 \rangle \simeq \langle \Delta \ell_{\rm \parallel, S}^2 \rangle > \langle \Delta \ell_{\rm \parallel, A}^2 \rangle$ for the initially scattering particles during the whole period. However, for the initially mirroring particles, we see $\langle \Delta \ell_{\rm \parallel, F}^2 \rangle \simeq \langle \Delta \ell_{\rm \parallel, S}^2 \rangle < \langle \Delta \ell_{\rm \parallel, A}^2 \rangle$ at $\tau \lesssim 6$, and $\langle \Delta \ell_{\rm \parallel, F}^2 \rangle \simeq \langle \Delta \ell_{\rm \parallel, S}^2 \rangle > \langle \Delta \ell_{\rm \parallel, A}^2 \rangle$ at $\tau \gtrsim 6$. In the case of perpendicular diffusion (panel d), we can see that the diffusion of initially mirroring particles satisfies $\langle \Delta \ell_{\rm \perp, A}^2 \rangle < \langle \Delta \ell_{\rm \perp, S}^2 \rangle < \langle \Delta \ell_{\rm \perp, F}^2 \rangle $ at $\tau \lesssim 10 $ and $\langle \Delta \ell_{\rm \perp, A}^2 \rangle > \langle \Delta \ell_{\rm \perp, S}^2 \rangle \simeq \langle \Delta \ell_{\rm \perp, F}^2 \rangle $ at $\tau \gtrsim 10$. As for the initially scattering particles, we can see $\langle \Delta \ell_{\rm \perp, A}^2 \rangle > \langle \Delta \ell_{\rm \perp, S}^2 \rangle \simeq \langle \Delta \ell_{\rm \perp, F}^2 \rangle $ during the whole periods. 

In short, the magnetosonic and the Alfv\'en modes dominate the parallel and perpendicular diffusion of the mirroring and scattering particles for a long evolution period, respectively. In the magnetosonic modes, the parallel diffusion of particles is faster. Whereas, the perpendicular diffusion of particles is faster in the Alfv\'en mode. Parallel diffusion is associated with the compressibility of the slow and fast modes, while perpendicular diffusion is associated with the random walk of the magnetic field dominated by the Alfv\'en mode.

\section{Discussion} \label{sec: style5}
In this work, we numerically explored the mirror and scattering diffusion of CRs in sub-Alfv\'enic and super-Alfv\'enic turbulence regimes first. After particles reached normal diffusion within several thousands of gyro periods in MHD turbulence, we measured the parallel and perpendicular MFPs of the CRs and quantified the relationships between MFPs and the Larmor radii. We found that the particles with larger initial pitch angles ($\mu_0 \in [0.1,0.2]$) diffuse more slowly than the particles with smaller initial pitch angles ($\mu_0 \in [0.8,0.9]$), which is consistent with \cite{Barreto-Mota2024arXiv}. The MFP of CRs is independent of the initial pitch angles. After testing the influence of local and global reference frames on the measurement results, we found that, for the power-law stages, the change in the frame of reference does not affect the power-law relation we obtained.

As predicted theoretically in \cite{Lazarian2014ApJ}, the CR perpendicular superdiffusion with the relation of $\langle \Delta \ell_{\perp} \rangle \propto t^{3/2}$ originates from the superdiffusion properties of turbulent magnetic field lines in the local frame of reference. Numerically, the perpendicular superdiffusion has been verified by \cite{Xu2013}, \cite{Hu2022MNRAS}, and \cite{Maiti2022ApJ}. Note that \cite{Maiti2022ApJ} explored CR superdiffusion behavior in the global and local references and found that the index of CR superdiffusion depends slightly on the choice of the frame of reference. Given that one cannot observe CR diffusion behavior in the local frame of reference, all results presented in this work are based only on the global frame of reference. In general, the efficiency of resonant scattering decreases for the same $M_{\rm A}$ with an increase in the inertial range. The results presented in this paper are limited to the numerical resolution of $512^3$ (see \citealt{Zhang2023ApJ} for a comparison between different resolutions).

Our current studies do not consider the CR acceleration and its radiative losses when the mirroring and scattering effects happen. \cite{CM2016A&A} has studied the propagation of CRs in MHD turbulence considering different forcing effects. They also measured the dependence of the MFP of CRs on their energy. For the sub-Alfv\'enic regime, they found the relations of $\lambda_{\rm \parallel} \propto R_{\rm g}^{1/3}$ for $M_{\rm A} >0.5$ and $\lambda_{\rm \perp} \propto R_{\rm g}^{0.67} $ for $M_{\mathrm A} =0.67$, which are consistent with the results presented in our paper (see the results for R2 and R3). Similarly, they also see inverted spectra with solenoidal forcing when $M_{\rm A} < 1$. In our work, we focus on the influence of different turbulence regimes on the propagation of CRs and measure MFPs of CRs within a wide energy range, i.e., from several grids to $L_{\rm inj}$. Considering the effect of acceleration on particles propagation, \cite{Beresnyak2011ApJ} found that the parallel diffusion coefficient is proportional to $R_{\rm g}$ for $R_{\rm g}<0.1L_{\rm inj}$ and $R_{\rm g}^2$ for $L_{\rm inj} > R_{\rm g} > 0.1L_{\rm inj}$ in trans-Alfv\'enic incompressible MHD turbulence. In addition, they also found that the perpendicular diffusion coefficient is independent of energy.

Our current work also explored the interaction of CRs with three MHD modes during the mirror and scattering diffusion. We found that the magnetosonic modes dominate parallel diffusion and the Alfv\'en mode dominates perpendicular diffusion. In our previous studies, we studied the CR acceleration, diffusion, and scattering of CRs due to the interactions of individual MHD modes. \cite{Zhang2021ApJ} found that the fast mode dominates the acceleration of particles in the case of super-Alfv\'enic and supersonic turbulence and the slow mode dominates the acceleration for sub-Alfv\'enic turbulence. As for the diffusion of the accelerated CRs, the slow mode dominates the diffusion of particles in the strong turbulence regime, whereas three modes have a comparable role in the weak turbulence regime (\citealt{Gao2024ApJ}). 

In addition to mirror diffusion explored in this paper, the understanding of the slow diffusion phenomena around the source region is attempted in other different ways including the anisotropy diffusion of CRs due to the anisotropy of MHD turbulence (e.g., \citealt{Liu2019PhRvL, Gao2025A&A}); the self-generated turbulence caused by CR flow restricting the diffusion of CRs (e.g., \citealt{Evoli2018PhRvD}); the constraint of diffusion from the strong turbulence generated by the shock wave of the parent supernova remnant (\citealt{Fang2019MNRAS}); and the two-zone mode assuming different diffusion coefficient near and far from the source region (\citealt{Fang2018ApJ}). In this paper, we confirmed that CRs undergo not only nonresonant interaction of magnetic mirrors but also gyroresonance interactions with the magnetic field during diffusion, exhibiting a synergistic effect of mirror and scattering diffusion \citep{Barreto-Mota2024arXiv}, which provides an alternative way to understand the diffusion process of CRs in different environments such as SNRs (see also \citealt{Xu2021ApJ} for application of mirror diffusion) and massive star-formation region.

\section{Summary} \label{sec: style6}
With the modern understanding of MHD turbulence, we performed test-particle simulations to study the mirror diffusion and scattering diffusion of CRs. The main findings of the paper are summarized as follows:

\begin{enumerate} 

\item Our simulations demonstrate that CRs with large pitch angles undergo mirror diffusion due to mirror reflection. In contrast, CRs with small pitch angles experience scattering diffusion due to gyroresonance with the magnetic field. The transition between the mirror and scattering diffusion appears when meeting or violating the following condition: the magnetic mirror size is larger than the Larmor radius of the particle with $\mu <\mu_c$ and $M=\rm const$. Comparing mirror diffusion with scattering one, the former has a stronger confining effect on CRs.

\item Regardless of the setting of the initial pitch angles, we find that CRs experience a transition from a superdiffusion to a normal diffusion after a sufficiently long evolution. During superdiffusion, the mean square displacement corresponding to the initially mirroring particles is significantly smaller than that to the initially scattering particles.

\item The normal diffusion stage is involved in a mixture of scattering and mirror diffusion, where the interaction of CRs with MHD turbulence results in a significant anisotropy of CR diffusion in the sub-Alfv\'enic regime. 

\item The CR diffusion strongly depends on the properties of MHD turbulence. We find the presence of the power-law relation between the mean free path and the Larmor radius $R_{\rm g}$. When $R_{\rm g}$ of CRs is in the range of strong turbulence, we have $\lambda_\perp \propto R_{\rm g}^{2/3}$ and $\lambda_\parallel \propto R_{\rm g}^{1/3}$ for the sub-Alfv\'enic regime as well as $\lambda_\perp \simeq \lambda_\parallel \propto R_{\rm g}$ for the super-Alfv\'enic regime. When $R_{\rm g}$ of CRs is in the range of weak turbulence ($M_{\rm A}<1$) or hydrodynamic turbulence ($M_{\rm A}>1$), we find the relation of $\lambda_\parallel \propto R_{\rm g}^2$, and the plateau-like distribution of $\lambda_\perp$.

\item Compressibility of the magnetosonic modes dominates the parallel diffusion of CR particles, while the random walk of the magnetic field associated with the Alfv\'en mode dominates the perpendicular diffusion of CR particles.

\end{enumerate}

\begin{acknowledgements}
We thank the anonymous referee for valuable comments that significantly improved the quality of the paper. We thank Jungyeon Cho for the helpful discussions on the referee's comments. Y.W.X. thanks Chao Zhang for the helpful discussions on numerical methods of test-particle simulations. J.F.Z. is grateful for the support from the National Natural Science Foundation of China (No. 12473046) and the Hunan Natural Science Foundation for Distinguished Young Scholars (No. 2023JJ10039). 
\end{acknowledgements}

\bibliographystyle{aa}
\bibliography{aa}
\end{document}